# Development of Multifractal Models for Self-Similar Traffic Flows

G. Millán*. G. Lefranc**

*Departamento de Ingeniería Eléctrica, Universidad de Santiago de Chile,
Santiago, Chile (e-mail: ginno.millan@usach.cl).
**Escuela de Ingeniería Eléctrica, Pontificia Universidad Católica de Valparaíso,
Valparaíso, Chile (e-mail: glefranc@ucv.cl).

**Abstract:** This paper presents a simple technique of multifractal traffic modeling. It proposes a method of fitting model to a given traffic trace. A comparison of simulation results obtained for an exemplary trace, multifractal model and Markov Modulated Poisson Process models has been performed.

*Keywords*: Computer networks, Fractals, Markov Modulated Poisson Process (MMPP), Multifractals, Self-similarity, Traffic modeling.

## 1. INTRODUCTION

Multifractal models (Millán, 2013) of network traffic use a unique mathematical construction called multifractal measure. These models, if properly constructed, can be useful when dealing with self-similar processes of network traffic intensity. The correct construction includes two fundamental elements: the model identification, and the model parameter estimation. This article proposes numerical methods to obtain an accurate multifractal traffic model. The first part an idea of multifractal models will be presented. In the second part are described two simple identification methods for multifractal models. In the third section, the method of model parameter estimation is presented. Using this model, a comparison of results is also obtained, including results for various Markov Modulated Poisson Process (MMPP) models.

## 2. BASIC IDEA OF MULTIFRACTAL MODELS

A general idea is derived from a construction of a multifractal measure special case: binomial measure (Millán and Lefranc, 2013).

The model consists of $N$ independent discrete time stochastic process $p_{i,j}$ where $i$ denotes a process index and $j$ denotes a process sample index. The model sample value with index $k$ can be obtained from

$$s_k = \prod_{i=1}^{N} p_{i,m}, \text{ with } m = \left\lfloor \frac{k}{2^{i-1}} \right\rfloor. \quad (1)$$

This means that a single process $p_2$ sample is used twice, a process $p_3$ sample is used 4 times, etc. The variance of a $p_i$ process determines the variance of a model at $2^i$ time scale.

## 3. IDENTIFICATION

The simplicity of the multifractal model structure implies a straightforward criterion for its identification. Since the $i$th process determines the self-similarity of the model at the $2^i$ time scale, the highest time scale $T$, at which the fractality is required, establish a minimum number of process equal to $\lceil \log_2 T \rceil$. However, in some cases this criterion fails, namely if characterized traffic trace is shorter than $T$. In this case, certainly, the trace is non-ergodic, and the mean value and variance of the traffic intensity cannot be estimated correctly. Multifractal properties of traffic flows make it possible to detect such a situation, so a non-ergodic trace can be detected and disqualified assuming it is fractal.

A right tool for this job is a multifractal analysis. In the real cases, the value of traffic intensity has its upper and lower bounds. Thus, assuming, that the model will need a finite number of processes (a model with an infinite number is non-realistic and could not be used for a simulation), each process will also have upper and lower bounds on its values. This conclusion limits a choice of process distribution to distributions defined in the interval. Among them, the Beta distribution appears to be the best choice, because it allows a continuous and widest range selection of the mean value and the coefficient of variance. The probability density function of Beta distribution is given by

$$f_\beta(x) = \begin{cases} 0 & , x \leq 0 \text{ and } x \geq 1 \\ \dfrac{x^{\alpha-1}(x-1)^{\beta-1}}{\beta(a,b)}, & 0 < x < 1 \end{cases}, \quad (2)$$

where $\beta(a, b)$ denotes the Beta function given by

$$\beta(a,b) = \int_0^1 x^{a-1}(1-x)^{b-1}dx, \ a,b \in \mathbb{R}^+. \quad (3)$$

In extreme cases this distribution reduces to either two-point distribution (a maximum of the coefficient of the variance) or a deterministic distribution (a minimum of the coefficient of the variance).

The simplest case is that all process $p_i$ are described by the same distribution (in this case Hurst exponent of the model process is equal to 1). When process $p_i$ is described by Beta distribution, its parameters should be positive. The analysis of this case leads to the conclusion that number of processes

$N$ should meet the following condition

$$N > \frac{\log[E(x)+\sigma^2(x)]-\log E(x)}{\log(\alpha_{max}+\alpha_{min}-\alpha_{max}\alpha_{min})}, \quad (4)$$

where $E(x)$ and $\sigma^2(x)$ denote a computed mean value and the variance of analysed trace; $\alpha_{min}$ and $\alpha_{max}$ denote minimum and maximum value of Hölder exponent observed in the trace respectively. If a value $2^N$ is greater or close to trace length, the trace should be treated a non-ergodic one.

The values $\alpha_{min}$ and $\alpha_{max}$ are obtained with use multifractal analysis of traffic pattern. The value of Hölder exponent $\alpha(x)$ in point $x$ for a given multifractal measure $\mu(x)$ is given by

$$\alpha(x) = \lim_{\delta \to 0} \frac{\log \sum_i \mu_i}{\log \delta}, \quad (5)$$

where the sum is computed over all values of measure define in the neighbourhood of $x$ with radius $\delta$. To transform a given pattern in a multifractal measure a pattern as defined in the interval [0, 1] has to be considered and normalized its values.

The sole purpose of the use of the described above Beta distribution is to obtain the identification criteria. Thus Beta distribution should not be considered as a requirement element of simulation model. However, the distribution of process $p_i$ in simulation model should be defined only for positive numbers since the model value is the multiplication of process values and the probability of a negative process value. A good choice is a logarithmic-normal distribution since there are accurate and efficient methods to generate pseudo-random numbers with this distribution.

## 4. MODEL PARAMETER ESTIMATION

The model structure makes it quite simple to determine parameters of the model. Each process $p_i$ viewed separately as in Fig. 1, for all time scales $T < 2^i$ is described by a constant value of variance.

Assume that a self-similar logarithmic-normal model process $s_i$ is a multiplication of several logarithmic-normal processes $p_{k,j}$. The process $p_k$ is defined in (1).

The self-similar process has various properties (Cox, 1984), which allow to identify and construct them in several ways. One of these properties is that a process variance decays according to the expression

$$\sigma^2(s_{i(T)}) \sim cT^{-\beta}, \quad (6)$$

where $s_{i(T)}$ denotes a process $s_i$ observed at a time scale $T$, $c$ is a constant scaling factor, and the shape parameter, $\beta$, can be expressed as follows

$$\beta = 2(H-1), \quad (7)$$

where $H$ is a Hurst exponent. The Process $s_{i(T)}$ is defined as

$$s_{i(T)} = \frac{1}{T}\sum_{m=1}^{T} s_{T(i-1)+m}. \quad (8)$$

The mentioned property allows obtain formulae determining parameters of process $p_{k,j}$. Compute a variance of process $s_i$

$$\sigma^2(s_i) = \sigma^2\left(\prod_{k=1}^{N} p_{k,j}\right) = E\left(\prod_{k=1}^{N} p_{k,j}^2\right) - E\left(\prod_{k=1}^{N} p_{k,j}\right)$$
$$= \prod_k E(p_{k,j}^2) - \prod_k E(p_{k,j})$$
$$= \prod_{k=1}^{N}[\sigma^2(p_{k,j})+E^2(p_{k,j})] - \prod_{k=1}^{N} E^2(p_{k,j}). \quad (9)$$

Since a mean value of process $s_i$ and process $s_{i(T)}$ are the same it is easier to perform the following computations with the use of the coefficients of variance instead of variances. Thus,

$$C^2(s_i) = C^2\left(\prod_{k=1}^{N} p_{k,j}\right) = \frac{\sigma^2\left(\prod_{k=1}^{N} p_{k,j}\right)}{E\left(\prod_{k=1}^{N} p_{k,j}\right)}$$
$$= \prod_{k=1}^{N}[C^2(p_{k,j})+1]+1. \quad (10)$$

Let $\sigma_i$ denote the coefficient of variance decay for time scale $2^i$, $\sigma_i = 2^{-2i(1-H)}$.

Combining (1)-(7) and (8), and assuming that

$$E(p_{k,j}) = \sqrt[N]{E(s_i)}, \quad (11)$$

for all $k$ we obtain the following system of $N$ equations

$$\begin{cases} C^2[s_i]\sigma_0 + 1 = \prod_{k=1}^{N}[C^2(p_{k,j})+1] \\ C^2[s_i]\sigma_1 + 1 = (2^{-1}C^2(p_{1,j})+1)\prod_{k=1}^{N}[C^2(p_{k,j})+1] \\ \vdots \\ C^2[s_1]\sigma_i + 1 = \prod_{k=1}^{i}(2^{k-i-1}C^2(p_{k,j})+1)\prod_{k=1}^{N}[C^2(p_{k,j})+1] \\ \vdots \\ C^2[s_1]\sigma_{N-1} + 1 = \prod_{k=1}^{N-1}(2^{k-1}C^2(p_{k,j})+1)(C^2(p_{N,j})+1) \end{cases} \quad (12)$$

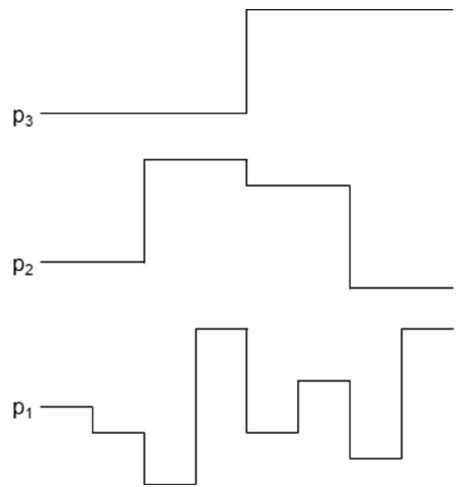

Fig. 1. Construction of multifractal model.

The system of equations (12) is illustrated with a graph in Fig. 2. The graph explains how the variance of model is constructed from the variances of processes $p_{k,j}$. The iterative solution of this system for values of $C^2(p_{k,j})$ obtained whit Gaussian elimination method can be written as follows

$$\begin{cases} L_1 = \sigma_0 C^2(s_i) + 1 \\ L_2 = 2\sigma_1 C^2(s_i) + 2 - L_1 \\ L_{n+1} = 1 \\ L_i = \dfrac{2\sigma_{i-1} C^2(s_i) + 2}{\prod_{k=1}^{i-2}[2^{k-i} C^2(p_{k,j}) + 1]} - L_{i-1} \\ C^2(p_{k,j}) + 1 = \dfrac{L_i}{L_{i+1}} \end{cases} \quad (13)$$

## 5. SIMULATIONS AND RESULTS

The purpose of performed simulations is check the behaviour of the presented model in application to network performance evaluation and compare a possible error of results obtained with the use of this model and other classic models. There exists a wide variety of networks interfaces which could be used in the reality to regulate a transmission of MPEG traffic. In this paper a simple model of Gigabit Ethernet IEEE 802.3ab is used. The model consist of an input network queue of Ethernet packets with length limited to 10000 places, a simple queuing model of Ethernet non-blocking switch; both input and output queues are limited to 5000 Ethernet packets per queue and sources of an additional traffic (Dong et al., 2010).

The compared results consisted of mean queue lengths at the network input switch, loss probabilities at shaper and switches and a variance of an inter-arrival time measured between the ends of subsequent MPEG frames on the output network.

The MPEG traffic model presented in this research has been compared with other MPEG traffic models: histogram based 30-states MMPP model and scene-oriented 300 states MMPP model (Ghandali and Safavi, 2011). Results obtained with the use of the original Star Wars sequence are assumed as a main reference point (Millán, 2018). Most simulation results (included mean queue lengths, loss probabilities) indicated no significant difference between multifractal, scene-oriented MMPP model and modelled trace. The results obtained with the use of histogram based MMPP were completely inaccurate. Larger differences between these models are visible in Fig. 3 which presents a variance of an inter-arrival time between subsequent MPEG frames measured on the network output. In this case none of compared models can be considered as a good approximation although the presented multifractal model turns up to be the best of them.

## 6. CONCLUSIONS

Presented method is an efficient way to simulate self-similar processes suitable especially for network traffic modelling since a model processes have positive values and can be constrained also a limited range without the loss of self-similarity and the change of generated process mean value and variance.

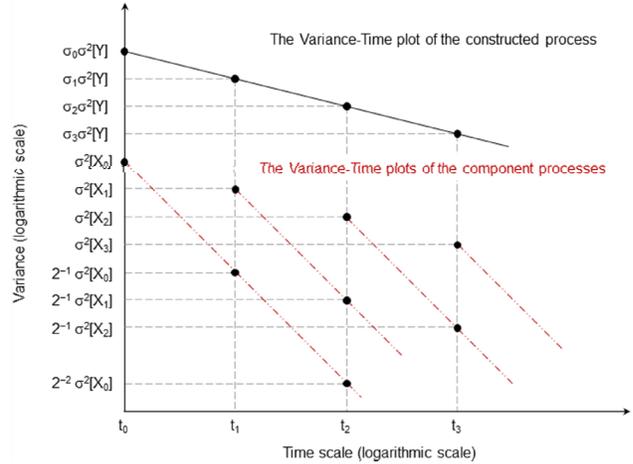

Fig. 2. Illustration of system equation given by (12).

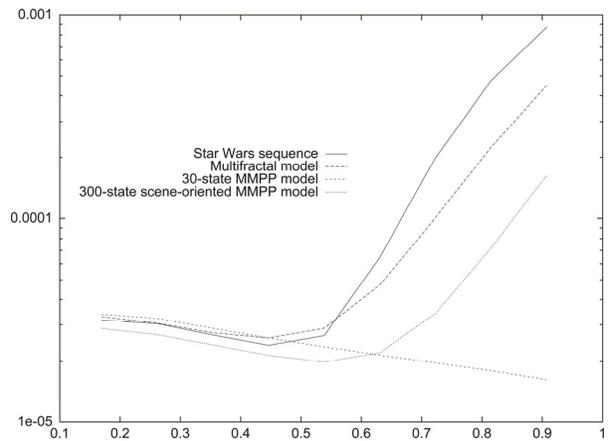

Fig. 3. A dependence of the MPEG frame inter-arrival time variance measured on the output of the network on the link load coefficient; X-axis: the link load coefficient, Y-axis: the inter-arrival time variance (square seconds).

ACKNOWLEDGMENT

Special thanks to the Advanced Human Capital Training Program of CONICYT Chile for the support given to the author G. Millán.